\documentclass[12pt]{iopart}

\begin{document}

\title{Geometric global quantum discord}

\author{Jianwei Xu}

\address{Key Laboratory for Radiation Physics and Technology, Institute of Nuclear Science and Technology, Sichuan University,
Chengdu 610065, China}
\ead{xxujianwei@yahoo.cn}
\begin{abstract}
Geometric quantum discord, proposed by Dakic, Vedral, and Brukner [Phys. Rev. Lett. 105 (2010) 190502], is an important measure for bipartite correlations. In this paper, we generalize it to multipartite states, we call the generalized version geometric global quantum discord (GGQD). We characterize GGQD in different ways, give a lower bound for GGQD, and  provide some special states which allow analytical GGQD.

\end{abstract}


\section{Introduction}

Quantum correlation is one of the most striking features in quantum theory.
Entanglement is by far the most famous and best studied kind of quantum correlation, and
leads to powerful applications \cite{Horodecki2009}. Another kind of quantum
correlation, called quantum discord, captures more correlations than
entanglement in the sense that separable states may also possess nonzero
quantum discord. Quantum discord has been attracted much attention in recent years, due
to its theoretical interest to quantum theory, and also due to its potential
applications \cite{Modi2011}.  Up to now, the studies on quantum correlations, like entanglement and quantum discord, are mainly focused on the bipartite case.

Quantifying the multipartite correlations is a fundamental and very intractable
question. The direct idea is that we can properly generalize the
quantifiers of bipartite correlations to the case of multipartite
correlations \cite{Coffman2000,Osborne2006,Zhou2006,Kaszlikowski2008,Giorgi2011}. Recently, generalizing the quantum discord of bipartite states to multipartite states
has been discussed in different ways \cite{Modi2011-2,Okrasa2011,Chakrabarty2010,Rulli2011}. As an important measure of bipartite correlations, the geometric quantum discord, proposed in \cite{Dakic2010}, has been extensively studied \cite{Modi2011}. In this paper, we generalize the
geometric quantum discord to multipartite states.

This paper is organized as follows. In Sec.2, we review the definition of geometric quantum discord for bipartite states. In Sec.3, we give the definition of geometric global quantum discord (GGQD) for multipartite states, and give two equivalent expressions for GGQD. In Sec.4, we provide a lower bound for GGQD by using the high order singular value decomposition of tensors. In Sec.5, we obtain the analytical expressions of GGQD for three classes of states. Sec.6 is a brief summary.

\section{Geometric quantum discord of bipartite states}

The original quantum discord was defined for bipartite systems over all
projective measurements performing only on one subsystem \cite{Zurek2001,Vedral2001}. That is, the quantum discord (with respect to $A$) of a bipartite state $\rho _{AB}$ of
the composite system $AB$ (we suppose $dimA=n_{A}<\infty ,$ $\dim
B=n_{B}<\infty $) was defined as
\begin{eqnarray}
D_{A}(\rho _{AB})=S(\rho _{A})-S(\rho _{AB})+\min_{\Pi _{A}}\{S(\Pi _{A}(\rho _{AB}))-S(\Pi _{A}(\rho _{A}))\},
\end{eqnarray}
In Eq.(1), $S(\cdot )$ is Von Neumann entropy, $\rho _{A}=tr_{B}\rho
_{AB}$, $\Pi _{A}$ is a projective measurement performing on $A$, $%
\Pi _{A}(\rho _{AB})$ is the abbreviation of $(\Pi _{A}\otimes
I_{B})(\rho _{AB})$ without any confusion, here $I_{B}$ is the
identity operator of system $B$. Note that $\Pi _{A}[tr_{B}(\rho _{AB%
})]=tr_{B}[\Pi _{A}(\rho _{AB})]$, that is, taking partial trace and
performing local projective measurement can exchange the ordering.

It can be proved that
\begin{eqnarray}
D_{A}(\rho _{AB})\geq 0,  \ \ \ \ \ \ \ \ \ \ \ \ \ \ \ \ \ \ \ \ \ \ \ \ \ \ \ \ \ \ \ \ \ \ \ \ \ \\
D_{A}(\rho _{AB})=0\Longleftrightarrow \rho _{AB}=\sum_{i=1}^{n_{A}}p_{i}|i\rangle \langle i|\otimes \rho _{i}^{B},
\end{eqnarray}
where, $n_{A}=\dim A$, $\{|i\rangle \}_{i=1}^{n_{A}}$ is any orthonormal
basis of system $A$, $\{\rho _{i}^{B}\}_{i=1}^{n_{A}}$ are density operators
of system $B$, $p_{i}\geq 0$, $\sum_{i=1}^{n_{A}}p_{i}=1$.

The original
definition of quantum discord in Eq.(1) is hard to calculate, even for
2-qubit case, by far we only know a small class of states which allow
analytical expressions \cite{Modi2011}.

 Dakic, Vedral, and Brukner proposed the
geometric quantum discord, as  \cite{Dakic2010}
\begin{eqnarray}
D_{A}^{G}(\rho _{AB})=\min_{\sigma _{AB}}\{tr[(\rho _{AB}-\sigma _{AB})^{2}]:D_{A}(\sigma _{AB})=0\}.
\end{eqnarray}
Obviously,
\begin{eqnarray}
D_{A}^{G}(\rho _{AB})=0\Longleftrightarrow D_{A}(\rho _{AB})=0.
\end{eqnarray}

For many cases $D_{A}^{G}(\rho _{AB})$ is more easy to calculate than $D_{A}(\rho _{AB})$
since $D_{A}^{G}(\rho _{AB})$ avoided the complicated entropy function. For instance, $%
D_{A}^{G}(\rho _{AB})$ allows analytical expressions for all
2-qubit states \cite{Dakic2010}, and also for all $2\times d$ $(2\leq d<\infty )$ states \cite{Rau2012}.

\section{Geometric global quantum discord}

In \cite{Rulli2011}, the authors generalized the original definition of quantum discord to
multipartite states, called global quantum discord (GQD).
Consider an $N$-partite $(N\geq 2)$ system, each subsystem $A_{k}$ $(1\leq
k\leq N)$ corresponds Hilbert space $H_{k}$ with dim$H_{k}=n_{k}$ (we
suppose $n_{k}<\infty $). The GQD of an $N$-partite state $\rho
_{A_{1}A_{2}...A_{N}}$ is defined as (here we use an equivalent expression
for GQD \cite{Xu2012})
\begin{eqnarray}
\fl
D(\rho _{A_{1}A_{2}...A_{N}})=\sum_{k=1}^{N}S(\rho _{A_{k}})-S(\rho
_{A_{1}A_{2}...A_{N}})   \nonumber  \\
-max_{\Pi }[\sum_{k=1}^{N}S(\Pi _{A_{k}}(\rho _{A_{k}}))-S(\Pi (\rho
_{A_{1}A_{2}...A_{N}}))],
\end{eqnarray}
where, $\Pi =\Pi _{A_{1}A_{2}...A_{N}}$ is a locally projective measurement
on $A_{1}A_{2}...A_{N}$.

Similar to Eqs.(2, 3), we have Lemma 1 below.

\emph{Lemma 1.}
\begin{eqnarray}
\fl
D(\rho _{A_{1}A_{2}...A_{N}})\geq 0,      \\
\fl
D(\rho _{A_{1}A_{2}...A_{N}})=0\Longleftrightarrow
\rho
_{A_{1}A_{2}...A_{N}}=
\sum_{i_{1}i_{2}...i_{N}}p_{i_{1}i_{2}...i_{N}}|i_{1}\rangle \langle
i_{1}|\otimes |i_{2}\rangle \langle i_{2}|\otimes ...\otimes |i_{N}\rangle
\langle i_{N}|.
\end{eqnarray}
Where, $\{|i_{k}\rangle \}_{i_{k}=1}^{n_{k}}$ is any orthonormal basis of $%
H_{k}$, $k=1,2,...N$, $p_{i_{1}i_{2}...i_{N}}\geq 0$, $%
\sum_{i_{1}i_{2}...i_{N}}p_{i_{1}i_{2}...i_{N}}=1$.

\emph{Proof.} Eq.(7) is proved in \cite{Rulli2011}. Eq.(8) can be proved as follows. Noting that $%
\Pi _{A_{1}A_{2}...A_{N}}(\rho _{A_{1}A_{2}...A_{N}})=\Pi _{A_{1}}(\Pi
_{A_{2}}...(\Pi _{A_{N}}(\rho _{A_{1}A_{2}...A_{N}}))...))$, then by Eq.(3)
and induction, Eq.(8) can be proved.

With Lemma 1, in the same spirit of defining geometric quantum discord for
bipartite states in Eq.(4), we now define the GGQD below.

\emph{Definition 1.} The GGQD of state $\rho _{A_{1}A_{2}...A_{N}}$ is defined as
\begin{eqnarray}
\fl  \ \ \ \ \ \ \ \ \ \
D^{G}(\rho _{A_{1}A_{2}...A_{N}})=
\min_{\sigma _{A_{1}A_{2}...A_{N}}}
\{tr[\rho _{A_{1}A_{2}...A_{N}}-\sigma
_{A_{1}A_{2}...A_{N}}]^{2}:D(\sigma _{A_{1}A_{2}...A_{N}})=0\}.
\end{eqnarray}

With this definition, it is obvious that
\begin{eqnarray}
D^{G}(\rho _{A_{1}A_{2}...A_{N}})=0\Longleftrightarrow D(\rho
_{A_{1}A_{2}...A_{N}})=0.
\end{eqnarray}

In \cite{Luo2010}, two equivalent expressions for Eq.(4) were given (Theorem 1 and Theorem 2 in \cite{Luo2010}), and they are very useful for simplifying the calculation of Eq.(4) and yielding lower bound of Eq.(4) \cite{Luo2010,Rana2012,Hassan2012}. Inspired by this observation, we now derive the corresponding versions of these two equivalent expressions for GGQD. These are Theorem 1 and Theorem 2 below.

\emph{Theorem 1.}  $D^{G}(\rho _{A_{1}A_{2}...A_{N}})$ is defined as in Eq.(9), then
\begin{eqnarray}
D^{G}(\rho _{A_{1}A_{2}...A_{N}})=\min_{\Pi }\{tr[\rho
_{A_{1}A_{2}...A_{N}}-\Pi (\rho _{A_{1}A_{2}...A_{N}})]^{2}\}       \ \ \ \ \ \ \ \ \ \ \ \ \ \ \ \    \nonumber   \\    \ \ \ \ \ \ \ \ \ \ \ \ \ \ \ \ \ \ \ \
=tr[\rho _{A_{1}A_{2}...A_{N}}]^{2}-\max_{\Pi }\{tr[\Pi (\rho
_{A_{1}A_{2}...A_{N}})]^{2}\},
\end{eqnarray}
where, $\Pi $ is any locally projective measurement performing on $%
A_{1}A_{2}...A_{N}.$

\emph{Proof.} In Eq.(9), for any $\sigma _{A_{1}A_{2}...A_{N}}$ satisfying $D(\sigma
_{A_{1}A_{2}...A_{N}})=0$, $\sigma _{A_{1}A_{2}...A_{N}}$ can be expressed
in the form
\begin{eqnarray}
\rho_{A_{1}A_{2}...A_{N}}
=\sum_{i_{1}i_{2}...i_{N}}p_{i_{1}i_{1}...i_{N}}|i_{1}%
\rangle \langle i_{1}|\otimes |i_{2}\rangle \langle i_{2}|\otimes ...\otimes
|i_{N}\rangle \langle i_{N}|,
\end{eqnarray}
where, $\{|i_{k}\rangle \}_{i_{k}=1}^{n_{k}}$ is any orthonormal basis of $%
H_{k}$, $k=1,2,...N$. $p_{i_{1}i_{2}...i_{N}}\geq 0$, $%
\sum_{i_{1}i_{2}...i_{N}}p_{i_{1}i_{2}...i_{N}}=1$. We now expand $\rho
_{A_{1}A_{2}...A_{N}}$ by the bases $\{|i_{k}\rangle
\}_{i_{k}=1}^{n_{k}}=\{|j_{k}\rangle \}_{j_{k}=1}^{n_{k}}$, $k=1,2,...N$.
Then
\begin{eqnarray}
\fl
\rho_{A_{1}A_{2}...A_{N}}
=\sum_{i_{1}j_{1},i_{2}j_{2},...,i_{N}j_{N}}
\rho_{i_{1}j_{1},i_{2}j_{2},...,i_{N}j_{N}}|i_{1}\rangle \langle j_{1}|\otimes
|i_{2}\rangle \langle j_{2}|\otimes ...\otimes |i_{N}\rangle \langle j_{N}|, \\
\fl
tr[\rho _{A_{1}A_{2}...A_{N}}-\sigma _{A_{1}A_{2}...A_{N}}]^{2}
=tr[(\rho_{A_{1}A_{2}...A_{N}})^{2}]+%
\sum_{i_{1}i_{2}...i_{N}}(p_{i_{1}i_{2}...i_{N}})^{2}     \nonumber \\    \ \ \ \ \ \ \ \ \ \ \ \ \ \ \ \ \ \ \
-2\sum_{i_{1}i_{2}...i_{N}}\rho
_{i_{1}i_{1},i_{2}i_{2},...,i_{N}i_{N}}p_{i_{1}i_{2}...i_{N}}  \nonumber \\  \ \ \ \ \ \ \ \ \ \ \ \ \ \ \
=tr[(\rho _{A_{1}A_{2}...A_{N}})^{2}]+\sum_{i_{1}i_{2}...i_{N}}(\rho
_{i_{1}i_{1},i_{2}i_{2},...,i_{N}i_{N}}-p_{i_{1}i_{2}...i_{N}})^{2}  \nonumber \\   \ \ \ \ \ \ \ \ \ \ \ \ \ \ \ \ \ \ \
-\sum_{i_{1}i_{2}...i_{N}}(\rho _{i_{1}i_{1},i_{2}i_{2},...,i_{N}i_{N}})^{2}.
\end{eqnarray}
Hence, it is simple to see that when $\rho
_{i_{1}i_{1},i_{2}i_{2},...,i_{N}i_{N}}=p_{i_{1}i_{2}...i_{N}}$ for all
$ i_{1},i_{2},...,i_{N}$, Eq.(14) achieves its minimum.

\emph{Theorem 2}. $D^{G}(\rho _{A_{1}A_{2}...A_{N}})$ is defined as in Eq.(9), then
\begin{eqnarray}  \fl  \ \ \ \ \ \ \ \ \
D^{G}(\rho _{A_{1}A_{2}...A_{N}})=\sum_{\alpha _{1}\alpha _{2}...\alpha
_{N}}(C_{\alpha _{1}\alpha _{2}...\alpha _{N}})^{2} \ \ \ \ \ \ \ \ \ \  \nonumber \\  \ \ \ \ \ \ \ \ \ \ \ \ \ \ \
-\max_{\Pi }\sum_{i_{1}i_{2}...i_{N}}(\sum_{\alpha _{1}\alpha _{2}...\alpha
_{N}}A_{\alpha _{1}i_{1}}A_{\alpha _{2}i_{2}}...A_{\alpha
_{N}i_{N}}C_{\alpha _{1}\alpha _{2}...\alpha _{N}})^{2},
\end{eqnarray}
where, $C_{i_{1}i_{2}...i_{N}}$ and $A_{\alpha _{k}i_{k}}$ are all real numbers, they are specified as
follows. For any $k$, $1\leq k\leq N$, let $L(H_{k})$ be the real Hilbert space
consisting of all Hermite operators on $H_{k}$, with the inner product $%
\langle X|X^{\prime }\rangle =tr(XX^{^{\prime }})$ for $X$, $X^{^{\prime
}}\in L(H_{k})$. For all $k$, for given orthonormal basis $\{X_{\alpha_{k}}\}_{\alpha_{k}=1}^{n_{k}^{2}}$ of $L(H_{k})$ (there indeed exists such a
basis, see \cite{Schlienz1995}) and orthonormal basis $\{|i_{k}\rangle \}_{i_{k}=1}^{n_{k}}$
of $H_{k}$, $C_{i_{1}i_{2}...i_{N}}$ and $A_{\alpha _{k}i_{k}}$ are
determined by
\begin{eqnarray}
\rho _{A_{1}A_{2}...A_{N}}=\sum_{\alpha _{1}\alpha _{2}...\alpha
_{N}}C_{\alpha _{1}\alpha _{2}...\alpha _{N}}X_{\alpha _{1}}\otimes
X_{\alpha _{2}}\otimes ...\otimes X_{\alpha _{N}},   \\
A_{\alpha _{k}i_{k}}=\langle i_{k}|X_{\alpha _{k}}|i_{k}\rangle.
\end{eqnarray}
\emph{Proof}. According to Eq.(11), and by Eqs.(16, 17), we have
\begin{eqnarray}  \fl  \ \ \ \ \ \
D^{G}(\rho _{A_{1}A_{2}...A_{N}})
=tr[\rho _{A_{1}A_{2}...A_{N}}]^{2}-\max_{\Pi }\{tr[\Pi (\rho
_{A_{1}A_{2}...A_{N}})]^{2}\}  \nonumber \\  \fl
=\sum_{\alpha _{1}\alpha _{2}...\alpha _{N}}(C_{\alpha _{1}\alpha
_{2}...\alpha _{N}})^{2}
-\max_{\Pi }\{tr[\sum_{i_{1}i_{2}...i_{N}}\sum_{\alpha _{1}\alpha
_{2}...\alpha _{N}}C_{\alpha _{1}\alpha _{2}...\alpha _{N}}\langle
i_{1}|X_{\alpha _{1}}|i_{1}\rangle \langle
i_{2}|X_{\alpha _{2}}|i_{2}\rangle \ \ \   \nonumber  \\    \ \ \ \ \ \ \ \ \ \ \ \ \ \ \ \ \ \ \ \ \ \
 ...\langle i_{N}|X_{\alpha
_{N}}|i_{N}\rangle |i_{1}\rangle \langle j_{1}|\otimes |i_{2}\rangle \langle
j_{2}|\otimes ...\otimes |i_{N}\rangle \langle j_{N}|]^{2}\} \nonumber  \\  \fl
=\sum_{\alpha _{1}\alpha _{2}...\alpha _{N}}(C_{\alpha _{1}\alpha
_{2}...\alpha _{N}})^{2}
-\max_{\Pi}\sum_{i_{1}i_{2}...i_{N}}(\sum_{\alpha
_{1}\alpha _{2}...\alpha _{N}}A_{\alpha _{1}i_{1}}A_{\alpha
_{2}i_{2}}...A_{\alpha _{N}i_{N}}C_{\alpha _{1}\alpha _{2}...\alpha
_{N}})^{2}.
\end{eqnarray}

\section{A lower bound of GGQD}

With the help of Theorem 2, we now provide a lower bound for GGQD.

If we regard $\rho _{A_{1}A_{2}...A_{N}}$ as a bipartite state in the
partition $\{A_{k},A_{1}A_{2}...A_{k-1}A_{k+1}...A_{N}\}$, then the original
quantum discord and geometric quantum discord of $\rho _{A_{1}A_{2}...A_{N}}$
with respect to the subsystem $A_{k}$ can be defined according to Eq.(1) and
Eq.(4), we denote them by $D_{A_{k}}(\rho _{A_{1}A_{2}...A_{N}})$ and $%
D_{A_{k}}^{G}(\rho _{A_{1}A_{2}...A_{N}})$. Comparing Eq.(3) and Eq.(8), it is
easy to find that
\begin{eqnarray}
D^{G}(\rho _{A_{1}A_{2}...A_{N}})=0\Longrightarrow D_{A_{k}}^{G}(\rho
_{A_{1}A_{2}...A_{N}})=0.
\end{eqnarray}
Consequently, comparing Eq.(4) and Eq.(9), we get
\begin{eqnarray}
D^{G}(\rho _{A_{1}A_{2}...A_{N}})\geq D_{A_{k}}^{G}(\rho
_{A_{1}A_{2}...A_{N}}).
\end{eqnarray}

To proceed further, we need a mathematical fact, called high order singular value
decomposition for tensors. We state it as Lemma 2.

\emph{Lemma 2}. \cite{Lathauwer2000} High order singular value
decomposition for tensors. For any tensor $T=\{T_{\beta _{1}\beta
_{2}...\beta _{N}}:\beta _{k}\in \{1,2,...,m_{k}\},k=1,2,...,N\},$
there exist unitary matrices $U^{(k)}=(U_{\beta _{k}\gamma _{k}})$, such that
\begin{eqnarray}
T_{\beta _{1}\beta _{2}...\beta _{N}}=\sum_{\gamma _{1}\gamma _{2}...\gamma
_{N}}U^{(1)}_{\beta _{1}\gamma _{1}}U^{(2)}_{\beta _{2}\gamma _{2}}...U^{(N)}_{\beta
_{N}\gamma _{N}}\Lambda _{\gamma _{1}\gamma _{2}...\gamma _{N}},\\
\sum_{\gamma _{1}\gamma _{2}...\gamma _{k-1}\gamma _{k+1}...\gamma
_{N}}\Lambda _{\gamma _{1}\gamma _{2}...\gamma _{k-1}\gamma _{k}\gamma
_{k+1}...\gamma _{N}}^{\ast }\Lambda _{\gamma _{1}\gamma _{2}...\gamma
_{k-1}\varepsilon _{k}\gamma _{k+1}...\gamma _{N}}=s_{\gamma
_{k}}^{(k)}\delta _{\gamma _{k}\varepsilon _{k}},     \\
s_{1}^{(k)}\geq s_{2}^{(k)}\geq ...\geq s_{n_{k}}^{(k)}\geq0.
\end{eqnarray}

Combining Lemma 2, Eq.(20) and the lower bound of $D_{A_{k}}^{G}(\rho
_{A_{1}A_{2}...A_{N}})$ in \cite{Luo2010}, we can readily obtain a lower bound of $%
D^{G}(\rho _{A_{1}A_{2}...A_{N}}).$

\emph{Theorem 3}. $D^{G}(\rho _{A_{1}A_{2}...A_{N}})$ is defined as in Eq.(9), then a lower bound of $D^{G}(\rho _{A_{1}A_{2}...A_{N}})$ is
\begin{eqnarray}
tr[(\rho _{A_{1}A_{2}...A_{N}})^{2}]-\min \{\sum_{\gamma
_{k}=1}^{n_{k}}s_{\gamma _{k}}^{(k)}:k=1,2,...,N\},
\end{eqnarray}
where $s_{\gamma _{k}}^{(k)}$ are obtained by Lemma 2 in which let $%
T=\{C_{\alpha _{1}\alpha _{2}...\alpha _{N}}:\alpha _{k}\in
\{1,2,...,n_{k}^{2}\},k=1,2,...,N\},$ $C_{\alpha _{1}\alpha _{2}...\alpha _{N}}$ are defined in Theorem 2.

\emph{Proof}. Since $D^{G}(\rho _{A_{1}A_{2}...A_{N}})$ and $D_{A_{i}}^{G}(\rho
_{A_{1}A_{2}...A_{N}})$ keep invariant under locally unitary transformation,
hence the state $\rho _{A_{1}A_{2}...A_{N}}$ in Eq.(16) and the state
\begin{eqnarray}
\Lambda _{A_{1}A_{2}...A_{N}}=\sum_{\alpha _{1}\alpha _{2}...\alpha
_{N}}\Lambda _{\alpha _{1}\alpha _{2}...\alpha _{N}}X_{\alpha _{1}}\otimes
X_{\alpha _{2}}\otimes ...\otimes X_{\alpha _{N}},
\end{eqnarray}
have the same GGQD, and
\begin{eqnarray}
D_{A_{k}}^{G}(\rho _{A_{1}A_{2}...A_{N}})=D_{A_{k}}^{G}(\Lambda
_{A_{1}A_{2}...A_{N}}).
\end{eqnarray}
From Eq.(20), we have
\begin{eqnarray}
D^{G}(\Lambda _{A_{1}A_{2}...A_{N}})\geq D_{A_{k}}^{G}(\Lambda
_{A_{1}A_{2}...A_{N}}).
\end{eqnarray}
From the definition of $D_{A_{k}}^{G}(\Lambda _{A_{1}A_{2}...A_{N}})$,
Lemma 2 and Theorem 1 in \cite{Luo2010}, we have
\begin{eqnarray}
D_{A_{k}}^{G}(\Lambda _{A_{1}A_{2}...A_{N}})=tr[(\rho
_{A_{1}A_{2}...A_{N}})^{2}]-\max_{\Pi _{A_{k}}}\sum_{i_{k}}(\sum_{\alpha _{1}\alpha _{2}...\alpha _{N}}A_{\alpha _{k}i_{k}}\Lambda _{\alpha _{1}\alpha _{2}...\alpha _{N}})^{2}   \nonumber \\
=tr[(\rho _{A_{1}A_{2}...A_{N}})^{2}]-\max_{\Pi _{A_{k}}}\sum_{i_{k}\alpha
_{k}}A_{\alpha _{k}i_{k}}^{2}s_{\alpha _{k}}^{(k)}   \nonumber \\
\geq tr[(\rho _{A_{1}A_{2}...A_{N}})^{2}]-\sum_{\alpha
_{k}=1}^{n_{k}}s_{\alpha _{k}}^{(k)}.
\end{eqnarray}
We then attain Theorem 3.

\section{Examples}

We provide some special states which possess analytical GGQD.

\emph{Example 1}. For $N$-qubit $(N\geq 2)$ Werner-GHZ state
\begin{eqnarray}
\rho =(1-\mu )\frac{I^{\otimes N}}{2^{N}}+\mu |\psi \rangle \langle \psi |,
\end{eqnarray}
the GGQD of $\rho $ is
\begin{eqnarray}
D^{G}(\rho )=\mu ^{2}/2.
\end{eqnarray}
In Eq.(29), $I$ is $2\times 2$ identity operator, $\mu \in \lbrack 0,1]$, $|\psi
\rangle $ is the $N$-qubit GHZ state
\begin{eqnarray}
|\psi \rangle =(|00...0\rangle +|11...1\rangle )/\sqrt{2}.
\end{eqnarray}
\emph{Proof}. We prove Eq.(30) according to Eq.(11).

$tr(\rho ^{2})$ can be directly calculated, that is
\begin{eqnarray}
tr(\rho ^{2})=(\frac{1-\mu }{2^{N}}+\mu )^{2}+(2^{N}-1)(\frac{1-\mu }{2^{N}}%
)^{2}.
\end{eqnarray}
$\max_{\Pi }\{tr[\Pi (\rho )]^{2}\}$ can be obtained by the similar
calculations of Theorem 4 In \cite{Xu2012}, the only difference is that the
monotonicity of entropy function under majorization relation (Lemma 4 in
\cite{Xu2012}) will be replaced by the case of the function
\begin{eqnarray}
f(p_{1},p_{2},...,p_{n})=-\sum_{i=1}^{n}p_{i}^{2}.
\end{eqnarray}
That is, $\max_{\Pi }\{tr[\Pi (\rho )]^{2}\}$ can be achieved by the eigenvalues
\begin{eqnarray}
\{\frac{1-\mu }{2^{N}}+\frac{\mu }{2},\frac{1-\mu }{2^{N}}+\frac{\mu }{2},%
\frac{1-\mu }{2^{N}},\frac{1-\mu }{2^{N}},...,\frac{1-\mu }{2^{N}}\}.
\end{eqnarray}
Thus
\begin{eqnarray}
\max_{\Pi }\{tr[\Pi (\rho )]^{2}\}
=2(\frac{1-\mu }{2^{N}}+\frac{\mu }{2})^{2}+(2^{N}-2)(\frac{1-\mu }{2^{N}}
)^{2}.
\end{eqnarray}
Combine Eqs.(32, 35), we then proved Eq.(30).

\emph{Example 2}. For $N$-qubit state
\begin{eqnarray}
\rho =\frac{1}{2^{N}}(I^{\otimes N}+c_{1}\sigma _{x}^{\otimes
N}+c_{2}\sigma _{y}^{\otimes N}+c_{3}\sigma _{z}^{\otimes N}),
\end{eqnarray}
the GGQD of $\rho $ is
\begin{eqnarray}
D^{G}(\rho )=\frac{c_{1}^{2}+c_{2}^{2}+c_{3}^{2}-\max
\{c_{1}^{2},c_{2}^{2},c_{3}^{2}\}}{2^{N}}.
\end{eqnarray}
In Eq.(36), $I$ is the $2\times 2$ identity operator, $\{c_{1},c_{2},c_{3}\}$
are real numbers constrained by the condition that the eigenvalues of $\rho $
must lie in $[0,1]$.

\emph{Proof}. We prove Eq.(37) by using Eq.(11).

$tr(\rho ^{2})$ can be directly found, that is
\begin{eqnarray}
tr(\rho ^{2})=\frac{c_{1}^{2}+c_{2}^{2}+c_{3}^{2}}{2^{N}}.
\end{eqnarray}
$\max_{\Pi }\{tr[\Pi (\rho )]^{2}\}$ can again be obtained similarly to Theorem 4
In \cite{Xu2012}, the only difference is that the monotonicity of entropy function
under majorization relation (Lemma 4 in \cite{Xu2012}) will be replaced by the case
of the function Eq.(33).

Similar reduction shows $\max_{\Pi }\{tr[\Pi (\rho )]^{2}\}$ can be achieved
by $\{\frac{1\pm c}{2^{N}}\}$, each of them have multiplicity $2^{N-1}$,
where $c=max\{|c_{1}|,|c_{2}|,|c_{3}|\}$. Therefore,
\begin{eqnarray}
\max_{\Pi }\{tr[\Pi (\rho )]^{2}\}=\frac{1+c^{2}}{2^{N}}.
\end{eqnarray}
Combine Eqs.(38, 39), we then get Eq.(37).

\emph{Example 3}. $N$-isotropic state
\begin{eqnarray}
\rho =(1-s)\frac{I^{\otimes N}}{d^{N}}+s|\phi \rangle \langle \phi |,
\end{eqnarray}
the GGQD of $\rho $ is
\begin{eqnarray}
D^{G}(\rho )=s^{2}(1-\frac{1}{d}),
\end{eqnarray}
where, $d=dimH_{1}$, $H_{1}=H_{2}=...=H_{N}$, $I$ is the $d\times d$
identity operator, $s\in \lbrack 0,1]$, $|\phi \rangle =\frac{1}{\sqrt{d}}%
\sum_{l=1}^{d}|ll...l\rangle $, $\{|l\rangle \}_{l=1}^{d}$ is an fixed orthonormal basis
of $H_{1}.$

\emph{Proof}. We prove Eq.(41) according to Eq.(9).
For any locally projective measurement $\Pi $, which corresponds $N$
orthonormal bases  of $H_{1}$, we denote them by $\{|i_{k}\rangle
\}_{i_{k}=1}^{d}$, $k=1,2,...,N$. Let $\{|l\rangle \}_{l=1}^{d}=\{|m\rangle
\}_{m=1}^{d}$. Then,
\begin{eqnarray}  \fl
\Pi (|\phi \rangle \langle \phi |)
=\frac{1}{d}\sum_{i_{1}i_{2}...i_{N},lm}\langle i_{1}|l\rangle \langle
m|i_{1}\rangle  ...
\langle i_{N}|l\rangle \langle m|i_{N}\rangle |i_{1}\rangle \langle
i_{1}|\otimes ...\otimes |i_{N}\rangle
\langle i_{N}|.     \\    \fl
tr\{[|\phi \rangle \langle \phi |-\Pi (|\phi \rangle \langle \phi |)]^{2}\}
=1-\frac{1}{d^{2}}\sum_{i_{1}i_{2}...i_{N}}(\sum_{lm}\langle i_{1}|l\rangle
\langle m|i_{1}\rangle...\langle i_{N}|l\rangle \langle m|i_{N}\rangle )^{2}   \nonumber   \\  \ \ \ \ \ \ \ \ \ \ \ \ \ \
\geq 1-\frac{1}{d^{2}}\sum_{i_{1}i_{2}...i_{N}}\sum_{lm}\langle
i_{1}|l\rangle \langle m|i_{1}\rangle...\langle i_{N}|l\rangle \langle m|i_{N}\rangle
=1-\frac{1}{d},
\end{eqnarray}
and the minimum can be achieved by taking  $\langle i_{k}|l\rangle =\delta
_{i_{k},l}$, $\langle m|i_{k}\rangle =\delta _{m,i_{k}}$. We then proved
Eq.(41).

We make some remarks. For states in Eq.(29) and states in Eq.(36), the GQD can also be analytically obtained \cite{Xu2012}, we then can compare the GQD and GGQD for these two classes of states. For  states in Eq.(36) and states in Eq.(40), when $N=2$, the GGQD in Eq.(37) and Eq.(41) recover the corresponding results  in \cite{Xu2012-2}.

We also remark that, from Eq.(37), let the state Eq.(36) undergo a locally phase channel performing on any qubit, similar discussions as in \cite{Cao2012} show that GGQD may also manifest the phenomena of sudden transition and freeze.

\section{Conclusion}

In summary, we generalized the geometric quantum discord of bipartite states
to multipartite states, we call it geometric global quantum discord (GGQD).
We gave different characterizations of GGQD which provided new insights for
calculating GGQD. As demonstrations, we provided a lower bound for GGQD by using the high order singular value decomposition of tensors, and obtained the analytical expressions of GGQD for three classes of
multipartite states. We also pointed out that GGQD can also manifest the phenomena of sudden transition and freeze.

Understanding and quantifying the multipartite correlations is a very
challenging question, we hope that the GGQD proposed in this paper may provide a useful attempt for this  issue.

\ack
This work was supported by the Fundamental Research Funds for the Central Universities of
China (Grant No.2010scu23002). The author thanks Qing Hou for helpful discussions.

\end{document}